\documentclass[fleqn,usenatbib]{mnras}

\usepackage{newtxtext,newtxmath}

\usepackage[T1]{fontenc}

\DeclareRobustCommand{\VAN}[3]{#2}
\let\VANthebibliography\thebibliography
\def\thebibliography{\DeclareRobustCommand{\VAN}[3]{##3}\VANthebibliography}

\usepackage{graphicx}	
\usepackage{amsmath}	

\title[HI Content of S0 Galaxies]{The Neutral Hydrogen Gas Content of Star-forming and Quiescent S0 Galaxies}

\author[D. A. A. Lee et al.]{
D. A. A. Lee$^{1}$, 
Karen L. Masters$^{2}$\thanks{E-mail: klmasters@haverford.edu}, David V. Stark$^{3,4}$ and Z. Z. Abidin$^{1,5}$\thanks{E-mail: zzaa@um.edu.my} 
\\
$^{1}$Radio Cosmology Research Lab, Centre for Astronomy and Astrophyiscs Research, Department of Physics, Faculty of Science, Universiti Malaya, \\ 50603, Kuala Lumpur, Malaysia
\\
$^{2}$Departments of Physics and Astronomy, Haverford College, 370 Lancaster Avenue, Haverford, PA 19041, USA\\
$^{3}$Space Telescope Science Institute, 3700 San Martin Drive, Baltimore, MD, 21218, USA\\
$^{4}$William H. Miller III Department of Physics and Astronomy, Johns Hopkins University, Baltimore, MD 21218, USA\\
$^{5}$National Centre for Particle Physics, Universiti Malaya, 50603 Kuala Lumpur, Malaysia\\
}

\date{Accepted 22 September 2026}

\pubyear{\the\year{}}

\begin{document}
\label{firstpage}
\pagerange{\pageref{firstpage}--\pageref{lastpage}}
\maketitle

\begin{abstract}

Lenticular (or S0) galaxies that remain actively star-forming (SFS0s) are a unique transition population among galaxies. We investigated the HI (21cm) neutral hydrogen content of a sample of lenticular galaxies identified from within the Mapping Nearby Galaxies at Apache Point Observatory (MaNGA) survey and which have been classified as star-forming or quiescent (qS0s). We compared these populations to each other and to a sample of star-forming spiral galaxies in MaNGA. We found that the HI Mass fraction and HI Deficiency of the SFS0s were intermediate between the quiescent S0s and the star-forming spirals, but most similar to the spirals. The HI depletion times of the quiescent S0's is very much longer than that of the SFS0 and spirals, which have roughly similar times. We also compared the molecular gas fractions of the three groups which shows that the lowest fraction of the three populations is found in the qS0s, followed by the SFS0s and then the SFSps.  A comparison of total baryonic mass indicates that the SFS0s have significantly lower masses than the qS0s in the sample. This suggests that the star-forming S0s must form from lower-mass progenitors (unobserved by MaNGA) which either accrete external gas, or loose spiral arms. The more massive quiescent S0s are a completely separate population, most likely to be formed either from fading star-forming spirals or from medium- to high-mass spirals undergoing mergers. The SFS0s in this sample cannot form from the quiescent S0s in this sample that receive in-falling gas, nor can they fade to create these qS0s.
\end{abstract}




\section{Introduction}
Lenticular galaxies, or S0 galaxies are disc galaxies lacking spiral features. The type was added to the Hubble classification scheme in \citet{sandage1961} between the elliptical and spiral types and they are often assumed to represent an intermediate stage in the evolution of galaxy morphology between the typically star-forming spiral galaxies and mostly quiescent ellipticals \citep[e.g.][]{KormendyBender1996ApJ,Salim2012,2025Privatus}. 

The likely formation pathways for S0s were recently reviewed in \citet[][]{Deeley2021MNRAS}, which used the IllustrisTNG simulations \citep[][]{Illustris2018MNRAS} to explore how S0s could potentially form. In that work, the authors suggested that, due to the diverse population, it is difficult to explain the formation of S0 galaxies with just one pathway. Other authors have also suggested S0s must have multiple formation paths including \citet{FraserMcKelvie2018,Xu2022MNRAS}. \citet{Deeley2021MNRAS} identified two potential formation pathways: gas stripping via group infalls (which they suggest make up 37\% of S0s) or formation via significant merger events (57\%).

\citet[][]{KoemendyKennicutt2004ARA&A} provide a useful review on how galaxy evolution in general takes place. They divide the possible mechanisms for galaxy evolution into those driven by external and internal factors (i.e. those depending on galaxy environment, and those more linked to internal process like AGN), and also separate the processes into either fast evolution (e.g. major mergers) or slow evolution (e.g. more gentle mass accretion, or internal orbital rearrangements). 
It has been known for a long time \citep[][]{dressler1980galaxy} that early-type galaxies (ETGs; usually defined as including both ellipticals, S0s although sometimes ETG can include spirals with large bulges) tend to be found in high-density environments where a lot of environmentally related processes occur. \citet[][]{goto2003} and many others find a similar relation in larger samples.

Most lenticular galaxies are quiescent \citep[i.e. not forming stars, e.g.][]{Stanford1998,Salim2012,GortTousSolanes2025}.  This fits into the commonly accepted view of galaxy evolution, where the S0 type is seen as a transition stage from star forming spirals to quiescent ellipticals \citep[e.g.][]{Bekki2002}. However quiescent spiral galaxies are known to exist \citep[e.g.][]{Masters2010,Cortese2012}, and blue early type galaxies \citep[e.g.][]{Lee2006ApJ,Schawinski2009MNRAS} present another interesting subset. 
Some-lenticular galaxies, are indeed observed to be star-forming \citep[e.g.][]{Rathore2022,Xu2022MNRAS}, and like quiescent spirals, these unusual star-forming S0s (SFS0s) provide an interesting sample in which we can attempt to disentangle the processes of galaxy evolution that may alter star formation properties and morphology separately. 

Previous work looking at star forming ETGs (either E+S0, or just S0s) include \citet[][]{Lee2006ApJ,Kannappan2009AJ, Schawinski2009MNRAS, Xu2022MNRAS, Rathore2022}. These works continue the overall picture of environment and mergers being important for the formation pathways of SFS0s. For example, in their study of 58 blue (i.e. star forming) ETGs, \citet[][]{Lee2006ApJ} found that almost 50\% showed signs of tidal disturbances. \citet{Schawinski2009MNRAS} reported on a population of blue ETGs from Galaxy Zoo, finding they tended to be found in lower-density environments than the red ETGs, but higher densities than typical spirals. \citet[][]{Rathore2022} in their investigation of SFS0 galaxies conclude they are mainly a result of minor mergers, with their sample of 120 SFS0s showing many signs of recent galaxy interactions such as kinematic misalignment, counter-rotation, and unsettled kinematics. Similarly \citet[][]{Xu2022MNRAS} identified minor mergers as important for SFS0 formation. In that work they compared SFS0 galaxies in the field to those in groups, finding that field SFS0s show spin parameters smaller than those of their group counterparts, suggesting different dynamical processes in their formation. It is clear that local environment plays an important role in the formation of SFS0s and blue ETGs. 

The radial extent of SF can provide useful information to constrain formation processes. In some samples of blue ETGs, extended SF seems common. For example \citet[][]{Kannappan2009AJ} present a study of nearby blue ETG and blue S0 galaxies. They note that the overall colors of these galaxies are driven by the presence of blue outer disks, suggesting inside out quenching dominates. They also comment that blue ETG sample were more likely to inverted colour gradients (i.e. blue inner and red outer) compared to red ETGs. \citet{Schawinski2009MNRAS} also noted spatially extended (and intense) SF in their sample and \citet{Deeley2021MNRAS} found in simulations that S0s forming via mergers tend to feature a transient star-forming ring. In contrast to these findings, \citet{Rathore2022} found more centrally concentrated SF in their sample of SFS0s (compared to SF spirals). This again suggests the presence of of multiple pathways for formation. 

Neutral atomic hydrogen (HI) provides a useful tracer for studying galactic evolution, that can be used to reveal the fuel for future star formation in a galaxy. Stars are formed from molecular hydrogen, not from HI directly and it is the surface density of molecular hydrogen that is best correlated with the current star formation rate \citep[SFR; ][]{Feldmann2020CmPhyLinkstargas}, while HI correlates best with sustained star formation \citep{Kannappan2013ApJ,Stark_2021}. But, since molecular hydrogen is formed out of HI, the study of HI is a useful measure of it's future potential in star formation. \citet{Rathore2022} note that molecular and/or atomic gas observations of their sample of SFS0s would be interesting to constrain formation timescales. For a long time, it was assumed S0s had no HI, however the work of \citet{Wardle1986} demonstrated HI content in a sample of S0 and Sa galaxies being intermediate between spirals and ellipticals. \citet[][]{vanDriel1991A&A} in summary of a series of observational papers, report on a sample of 25 early-type disk galaxies (S0 to Sa) observed in HI. At that time the discovery of HI in such early-type discs challenged theories of galaxy formation. \citet{vanDriel1991A&A} suggest that either recent accretion from other galaxies, or a burnt out disk scenario could explain the result.  More recent work investigating HI in early type galaxies includes \citet[][]{SerraAtlas3D2012MNRAS} who observe 166 ETG in the Atlas3D sample using the Westerbork Synthesis Radio Telescope. Their findings point towards the trend of non-cluster early types to hold on to their HI content, with the higher probability of said HI being in a settled configuration. In further work with a similar sample \citet{Kokusho2017} considered the SF properties of these ETGs, concluding that lower SFR in ETGs was not typically due to lower SF efficiency than spirals, but rather reduced gas content, particularly molecular gas.

Reviewing other work considering molecular hydrogen, in S0s, the general picture is that these types of galaxies are not simply passive, gas-poor systems. Their molecular gas content is lower than spirals as expected. However, when gas is present—most likely acquired externally—it can fuel efficient, possibly bursty star formation. For example, \citet[][]{Sage2007ApJ} looked at the molecular gas in a volume-limited sample of 46 non-cluster elliptical galaxies finding that the cool molecular gas in ellipticals tend to show the mass of cool gas in S0 galaxies cuts off at ~10 percent of what is expected from current models of gas return from stellar evolution. They also found lower detection rates and possibly much lower H$_2$/HI mass ratios in the ellipticals. The detection rate is higher among the lower mass galaxies, which they found strange as the higher mass galaxies tend to have greater gravitational potential wells. 
Similarly, \citet[][]{Wei2010ApJb} looked at molecular hydrogen (H$_{2}$) in 19 low mass early type galaxies, with emphasis on those in the blue cloud. The authors note that 80$\%$ of their blue samples exhibit apparently higher molecular gas star formation efficiency than the local spirals. The possible explanations outlined in this work indicates either a star-formation that occurs in bursts or the failure of the CO(1-0) to fully trace the molecular hydrogen. The fact that the authors suggested that the star formation occurs in bursts may point towards these galaxies getting their molecular gas supplies replenished by possible accretion events.

The main motivation for this work can be laid out as follows. To form stars, any galaxy has to have a reserve of gas available. The cold gas in the interstellar medium is mostly HI, by mass \citep{Saintonge2022}. Using data from the HI-MaNGA survey \citep{Masters_2019}, we will quantify the amount of neutral atomic hydrogen gas available in a sample of star-forming lenticular galaxies. We will compare the HI content, deficiency, and depletion times for the star-forming lenticulars to control samples of both quiescent lenticulars and star-forming spirals to see how HI content can help inform the likely formation scenario for SFS0s. 
 
The paper is laid out as follows: in Section 2, we discuss our data sources and sample of the galaxies used in this work, including optical IFU data and sample selection (\S 2.1) and HI data (\S2.2); Section 3 shows our results which we discuss in context of other work on the HI content of SFS0s and formation scenarios for these objects in Section 4. We conclude in Section 5. 

Where physical units are noted, we assume a Hubble constant H$_0$= 70.0 km s$^{-1}$ Mpc$^{-1}$, $\Omega_A$ =0.69, and $\Omega_m$= 0.31. 

\section{Sample and Data Sources}
\subsection{Optical IFU Data and Sample Selection} \label{sec:sample}

This work is based on the sample of star-forming or quiescent lenticulars identified by \citet[][hereafter R22]{Rathore2022} which are a selected out of the sample of galaxies in from the Mapping Nearby Galaxies at Apache Point Observatory \citep[MaNGA, ][]{Bundy2015MaNGA} survey; an Integrated Field Unit (IFU) survey which provides resolved optical spectroscopy data for 10,001 nearby galaxies and was part of the fourth phase of the Sloan Digital Sky Survey \citep[SDSS-IV, ][]{Blanton2017}.

\subsubsection{Selection of SFS0 and Control Samples}
MaNGA targets were selected out of the NASA Sloan Atlas \citep[NSA, ][]{Blanton2011} reprocessing of SDSS legacy imaging \citep{York2000} to create a sample of galaxies with a flat stellar mass distribution across $\log(M_\star/M_\odot)= 9-12$ which have angular sizes which fit the distribution of the MaNGA IFU bundles \citep{Drory2015}. A primary sample, making up 80\% of the galaxies, are covered to a radius of $1.5r_e$ (where $r_e$ is the effective radius), while the secondary sample are covered to $2.5r_e$. For more details of MaNGA target selection, see \citet{Wake2017}. 

In order to select a sample of lenticular galaxies from MaNGA, R22 made use of the DR15 "deep learning" morphology catalog published in \citep{sanchez2018MNRAS}, which used Galaxy Zoo and other morphologies to train a CNN model\footnote{Galaxy Zoo morphologies are also directly available for almost all MaNGA galaxies at \url{https://www.sdss4.org/dr17/data_access/value-added-catalogs/?vac_id=galaxy-zoo-classifications-for-manga-galaxies}\citep{Abdurrouf2022}}. R22 made a pre-selection using $P_{S0}>0.5$ and {\tt TTYPE}$>0$ to identify candidate S0 galaxies, then carried out additional visual inspection based on the deepest imaging available for each galaxy to check that no faint spiral structure was present. To classify galaxies as star-forming or quiescent, R22 used estimates of star formation rates (SFR), and stellar masses of the galaxies provided by the GSWLC-A2 \citep[GALEX-SDSS-WISE-LEGACY Catalogue, ][aka the "Salim catalog"]{Salim2016ApJ,Salim2018ApJ}. 

To reconstruct a volume-limited sample, we make use of the sample weights provided by \citet{Wake2017} for all MaNGA target galaxies. We follow \citet[][priv. comm.]{Rathore2022}, and use the weights for the combined primary+ and full secondary sample, labeled as '{\tt esweight}' in the MaNGA DRP data model. This gives the widest sample to draw from.  
Note, however, that 1 to 3 percent of the galaxies in the R22 sample are drawn from MaNGA ancillary samples, and hence have no weights (all sample sizes are summarized in Table \ref{tab:Table_1}). We find that the inclusion or not of these weights has minimal impact on the distribution of properties.

 \begin{table*}
	\centering
	\caption{A summary of the sample size of the different galaxy samples used in this work.}
	\label{tab:Table_1}
	\begin{tabular}{lcccccr} 
		\hline
		Category & R22 sample & Has {\tt esweight} & HI-MaNGA (DR4) & HI Detections & HI Upper limits \\
		\hline
		Quiescent S0 & 227 & 224 & 174 & 25 (14\%) & 149 (86\%) \\
		Star-Forming S0 & 120 & 117 & 109 & 28 (26\%) & 81 (74\%) \\
		Star Forming Spirals & 1468 & 1428 & 1291 & 889 (69\%) & 402 (31\%) \\
		\hline
	\end{tabular}
\end{table*}

\subsubsection{Measurements based on Optical Data}
For consistency with R22, we make use of the SFRs provided by \citet{Salim2016ApJ},  which are estimated by fitting models to the ultraviolet (UV), optical, and mid-infrared (IR) broad-band spectral energy distribution (SED) of galaxies. \citet{Salim2016ApJ} only provide SFR for galaxies that have a {\tt SED flag} = OK, which excludes galaxies that show broad emission lines, characteristic of broad-line AGNs. As broad band SED fits, these SFRs are representative of the SFR averaged over the last 100 Myrs for each galaxy. 

We make use of estimates of the total molecular gas mass from {\tt Pipe3D} \citep[][Pipe3D column {\tt log\_Mass\_gas\_Av\_gas\_log\_log} as recommended\footnote{Note that this {\tt Pipe3D} column differs from the {\tt log\_Mass\_gas} column which is a different estimate for the molecular gas}]{Sanchez2022}. Pipe3d is a set of data products extracted from the raw MaNGA data via the {\tt pyPIPE3D} pipeline \citep[][]{Sanchez2022}. These data products are generally the result of analysis of the stellar populations of the individual galaxies and/or ionized gas emission line fluxes and dynamics. Molecular gas masses are estimated in the Pipe3D analysis via a correlation with dust extinction measured using the Balmer decrement -- specifically, \citet{Sanchez2022} recommends using values from the calibration $\Sigma_{\rm mol} = 1.06 A_{V,{\rm gas}}^{2.58}$, where $\Sigma_{\rm mol}$ is the molecular gas surface density in solar masses per pc$^2$ and $A_{V,{\rm gas}}$ is the extinction in magnitudes (their Equation 10, which is based on a fit to EDGE-CALIFA data discussed in \citealt{Barrera-Ballesteros2020})\footnote{This calibration is discussed further in \citet{Barrera-Ballesteros2021manga} where they used the same data to measure molecular gas content via the correlation with CO emission and compare to the extinction estimates from the Balmer decrement, finding the very similar relation or $\Sigma_{\rm mol} = 10^{1.37} A_{V,{\rm gas}}^{2.3}$ (see their Section 3.1). The difference is attributed to slightly different sample selections and fitting techniques (Barrera-Ballesteros, priv. comm.)}. The Balmer decrement technique estimates dust extinction, $A_{V, {\rm gas}}$ assuming the ionized gas is at $T=10^4$K, which is a reasonable estimate for dust warmed by SF. This technique works because the dust-to-gas ratio in the interstellar medium is relativity constant, albeit metallicity dependent. This technique means that molecular gas mass estimates are only available for galaxies with emission line detections in the MaNGA data, and in addition only estimate molecular gas masses in the region covered by the MaNGA bundle. Surveys of the CO content to trace molecular gas (e.g. ALMA-MaNGA QUEnching and STar formation {\citep[ALMaQUEST, ][]{2020ApJalmaquest}, and the upcoming KILOGAS \citep[][]{Davis2026}  would present an alternative approach to estimate molecular hydrogen mass, they are are currently only available for a very small fraction of our sample, and we opt to use the Balmer-decrement approach to ensure self-consistent estimates of H$_{2}$ in our study. Recently \citet{Scholte2023} investigated the use of the Balmer decrement method for estimating molecular gas masses, concluding it is a promising technique to obtain estimates for larger samples with optical spectroscopy. 

Finally, we made use of stellar masses derived from  {\tt elpetro\_{mass}} from the NASA Sloan Atlas \citep[NSA; ][]{Blanton2011}. This is the stellar mass provided in HI-MaNGA file, and is adjusted to H$_0$=70 in HI-MaNGA (See Section 2.2). We have tested whether our results are dependent on the specific stellar mass estimates used by comparing the results using the stellar masses from the NSA mass, masses from Pipe3D and the SED fitting based mass from \citet{Salim2016ApJ} and found no significant qualitative differences in the results.

\subsection{Neutral Atomic Hydrogen (HI) Data}

A MaNGA follow-up program called HI-MaNGA \citep[][]{Masters_2019,Stark_2021} was initiated with the aim of complementing existing MaNGA data with HI observations, either from published results (from the Arecibo Legacy Fast ALFA Survey, or ALFALFA \citet[][]{Haynes_2018}) or with new observations carried out using the Robert C. Byrd Green Bank Telescope (GBT). In this work, we use data from the 4th data release (DR4)\footnote{\tt https://greenbankobservatory.org/science/gbt-surveys/hi-manga/} from HI-MaNGA which contains information on the HI content of 7013 MaNGA galaxies.
For details on the reduction in data from the GBT HI data, and homogenization of measurements of HI mass and line widths with published work, see \citet{Masters_2019,Goddy2020,Stark_2021}.\footnote{\tt https://www.sdss4.org/dr17/manga/hi-manga/}

The galaxy catalogs from \citet[][]{Rathore2022} were cross-matched with the HI-MaNGA catalog, based on the MaNGA plate-IFU number.  Our final samples are summarized in Table \ref{tab:Table_1} shows the number of galaxies in the original \citet{Rathore2022} sample, those with {\tt es weight}, those with HI-MaNGA DR4 observations, further separated into detections and upper limits. We show the SFR vs. stellar mass relations of the three samples in Figure \ref{fig:1}\footnote{For the coding for this Figure, and in other parts of the analysis, we make use of Google Collab with assistant from the integrated AI in Google Collab.}, which shows how the SF S0s and spirals populate the star-forming sequence, while the quiescent S0s are significantly below it. This plot also indicates whether galaxies have HI detections or upper limits.

HI-MaNGA provides data on both detections and upper limits of MaNGA galaxies. The HI Mass of the galaxies, $M_{\rm HI}$ is measured in HI-MaNGA detections using the standard equation (refer to \citet[][]{Stark_2021} for details).
Where HI-MaNGA does not detect a galaxy, upper limits are provided which estimate the maximum possible HI mass which could have remained undetected based on the depth of the observation. These use the measured $rms$ noise of the spectrum and assume a total line width of 200~km~s$^{-1}$.

We show in Figure \ref{fig:2} histograms of the HI mass  (detections or upper limits) for the three sub-samples of galaxies. This clearly shows that both S0 samples, but most particularly the quiescent S0 sample, are dominated by HI upper limits, while the SF spiral sample has more detections than upper limits. This large number of upper limits in our data, means that the statistical techniques of survival analysis will be important to ensure we capture the maximum information from our dataset. Survival analysis is an overarching term for a set of statistical techniques which can deal with upper (or lower) limits on data, in addition to known values.\footnote{The survival analysis techniques were originally developed to account for deaths of patients in the statistics of survival times after injury or disease, hence the name, but in astronomical use becomes useful to include non-detection (or upper limits) in statistics. We refer the reader to \citet{Stark_2021} for a more extended discussion of the use of such techniques with HI observations.}.

We calculate a HI deficiency factor, HI$_{\rm Def}$ for galaxies in the sample, using either the HI mass or the upper limit (where this becomes a lower limit on the true deficiency). This is calculated using the formula: 
 \begin{equation}
   HI_{\rm Def} = \langle \log(M_{\rm HI}/M_{\star})\rangle - \log(M_{\rm HI}/M_{\star}), 
	\label{eq:1}
\end{equation}
where $\langle f_{\rm HI}\rangle = \langle \log(M_{\rm HI}/M_{\star})\rangle = -0.786 \log(M_{\star}/M_\odot)+7.166$ is the survival fit to the HI mass fraction-stellar mass relation shown in Figure \ref{fig:3}. 

 This ``survival" fit refers to a technique for fitting straight lines which accounts for both detection and the upper limits, using the Akritas-Thiel-Sen \citep[ATS, ][]{Akritas1995} method which is known to be minimally biased to the point where the fraction of upper limits exceeds 50\% \citep[][]{Stark_2021}. The ATS method finds the best slope from a set of data including both detections and upper limits by determining the slope that yields a Kendall tau correlation coefficient of zero when subtracted from the data. The method uses a modified version of Kendall’s tau  (a type of rank correlation test) which considers upper limits when testing correlation strength, identifying them in concordant or discordant data point pairs when possible, or otherwise treating them as ties. The intercept is the midpoint of the cumulative distribution function (CDF) of the residuals after the best-fit slope is removed, derived using the Kaplan-Meier estimator. The Kaplan-Meier estimator is another statistical technique in the ``survival analysis" toolkit, which can generate a cumulative distribution which accounts for upper-limits. The Kaplan-Meier estimator can be used to calculate the cumulative probability of a data set with non-detections, where the probability distribution of each non-detection is effectively evenly re-distributed among all detections existing at or below that upper limit.  For a full description of the method and its application to astronomy, see \citep[][]{FeigelsonandNelson1985ApJ} and \citep[][]{FeigelsenandBabubook}. No weights were used for this particular part.

Using the star formation rate (SFR) provided in \citet[][]{Rathore2022} data, we also estimated the atomic gas depletion time, $\tau_{HI}$ in Gyrs as \begin{equation}
  \tau_{HI} = M_{HI} / SFR, 
	\label{eq:2}
\end{equation}
and equivalent upper limits on this where we have only upper-limits (or non-detections) for the HI content.
  
\begin{figure*}
	\centering
    \includegraphics[width=1.0\textwidth]{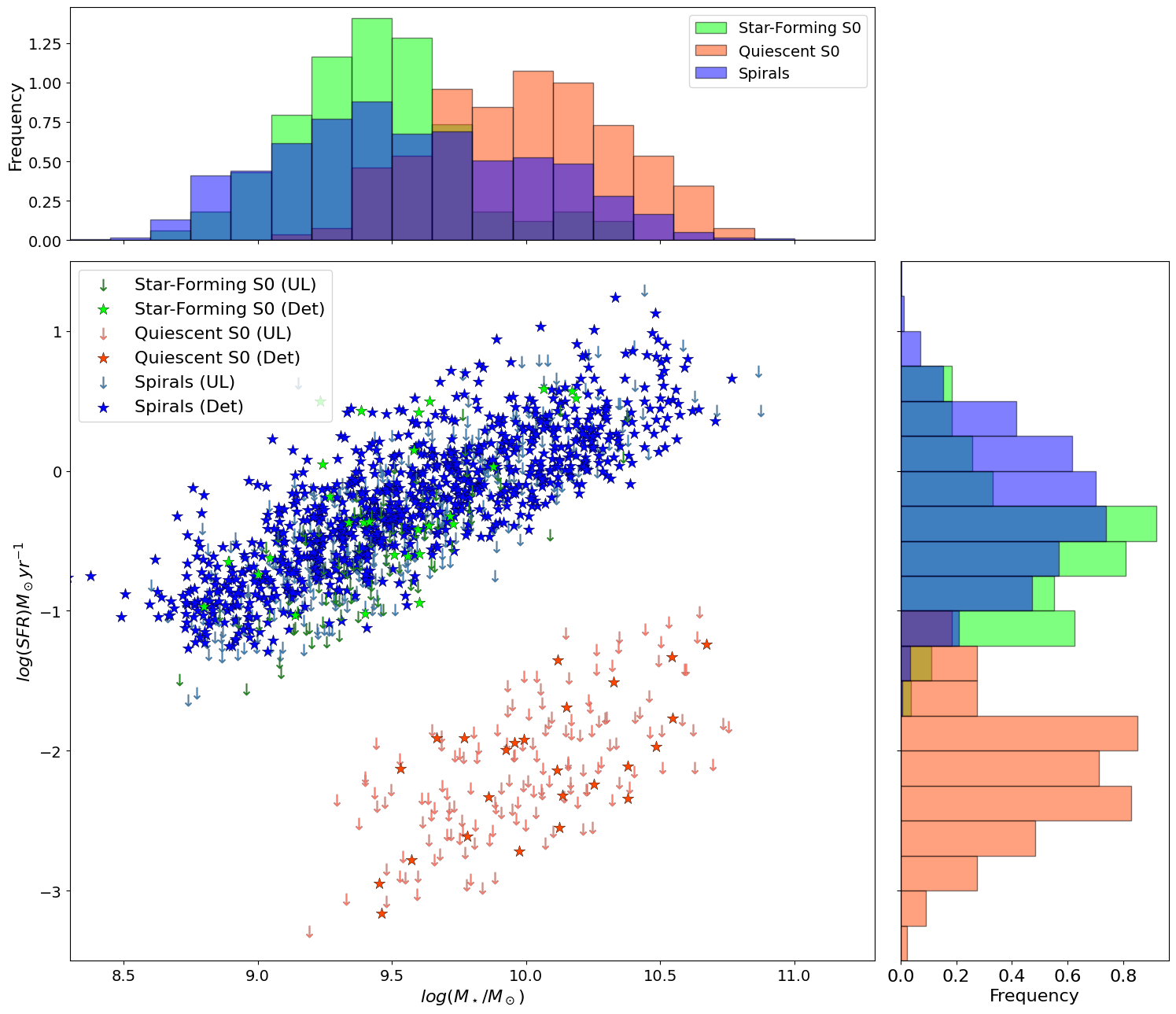}
    \caption{SFR in solar masses per year is plotted against total stellar mass for the galaxies considered in this paper. In this, and all following plots we show star forming spiral galaxies in blue, star forming lenticular galaxies in green and quiescent lenticulars in red/orange. Where galaxies are detected in HI-MaNGA they are plotted as a star; galaxies with HI upper limits are shown as downward arrows. Also included are two histograms, one showing the distribution od stellar mass for each category, while another shows the distribution of Star Formation Rate for the three categories.} 
    \label{fig:1}
\end{figure*}

\begin{figure*}
	\includegraphics[width=1.0\textwidth,]{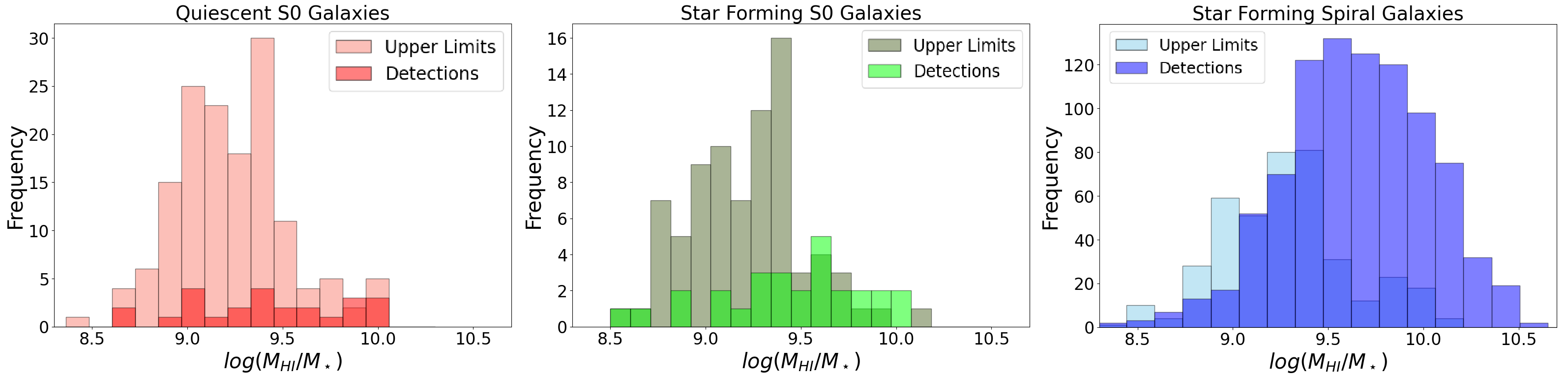}
    \caption{Histograms of the HI mass, or HI upper limit (as indicated in the legend) for (left panel) the quiescent S0 galaxies,(middle) the star-forming S0s; and (right) star-forming spirals from the sample of \citet[][]{Rathore2022}. For the Star-forming S0's the fractions of detections to upper limits are 28 detections (26\%) and 81 upper limits, the quiescent S0'S 25 detections (14\%) and 149 upper limits and the star-forming spirals 889 detections (69\%) and 402 upper limits.}
    \label{fig:2}
\end{figure*}

\subsubsection{Baryonic Mass}
Using the HI data in combination with optical measurements we are also able to estimate total baryonic masses for these galaxies. To calculate the total baryonic mass ($M_{\rm bary}$, we used the equation, 
\begin{equation}
M_{\rm bary} = 1.07M_{*} + \frac{4}{3}M_{\rm HI},
\end{equation}
where the 7\% addition to $M_\star$ is an estimate to account for molecular hydrogen, \citep[based on][]{McGaugh2020RNAAS}. In Section \ref{sec:molecular} we directly estimate molecular gas mass fractions for a subset of the sample, based on a correlaton with dust content. This 7\% addition is a good estimate on average for the quiescent S0's (qS0s), and a underestimate for our SF samples. The factor of $4/3$ on $M_{\rm HI}$ accounts for the cosmic abundance of helium atoms. 

\section{Results}

\subsection{Comparing HI Mass Fractions}
\label{sec:Mass Distribution} 

We show the HI mass distribution of our three samples in Figure \ref{fig:2} indicating detections and upper limits separately. Since HI masses scale with overall galaxy mass, we compare HI mass fractions, $\log(M_{\rm HI}/M_\star)$, which are shown for all three samples plotted against stellar mass in Figure \ref{fig:3}. As is clear from these figures there are substantial fractions of HI upper limits in both S0 samples (74\% and 86\% of the SF and quiescent subsets are non-detections in HI-MaNGA). In order to statistically compare samples with significant numbers of upper limits, we will make use of the CDF generated via the Kaplan-Meier ``survival analysis" technique$^6$ which accounts for both events (detections) and upper-limits in generating a CDF and confidence interval. From this we can then calculate the median and {\bf it's} confidence interval where the CDF crosses 0.5 (\citealt{KaplanMeier}; also see \citealt{Stark_2021} who previous use this technique with HI-MaNGA data).  

The CDFs of the HI mass fraction for our three subsets are shown in Figure \ref{fig:4}. It is clear from this figure that qS0s as a whole have the lowest HI mass fractions (median $\log(M_{\rm HI}/M_\star) = -0.6 ^{+0.8}_{-0.9}$, followed by SFS0s (median $\log(M_{\rm HI}/M_\star) = -0.06 ^{+0.2}_{-0.06}$ then spirals (median $\log(M_{\rm HI}/M_\star) = 0.3^{+0.04}_{-0.03}$. Comparing the distributions using the logrank test\footnote{From the Python module {\tt lifelines}; \\ {\tt https://lifelines.readthedocs.io/}} we exclude the possibility ($p<0.005$) that the SF and quiescent lenticular HI fractions are drawn from the same population, but find that it is likely ($p=0.2$) that the SFS0s have statistically the same HI mass fractions as the spiral sample. 

\begin{figure*}
	\centering
    \includegraphics[width=\textwidth]{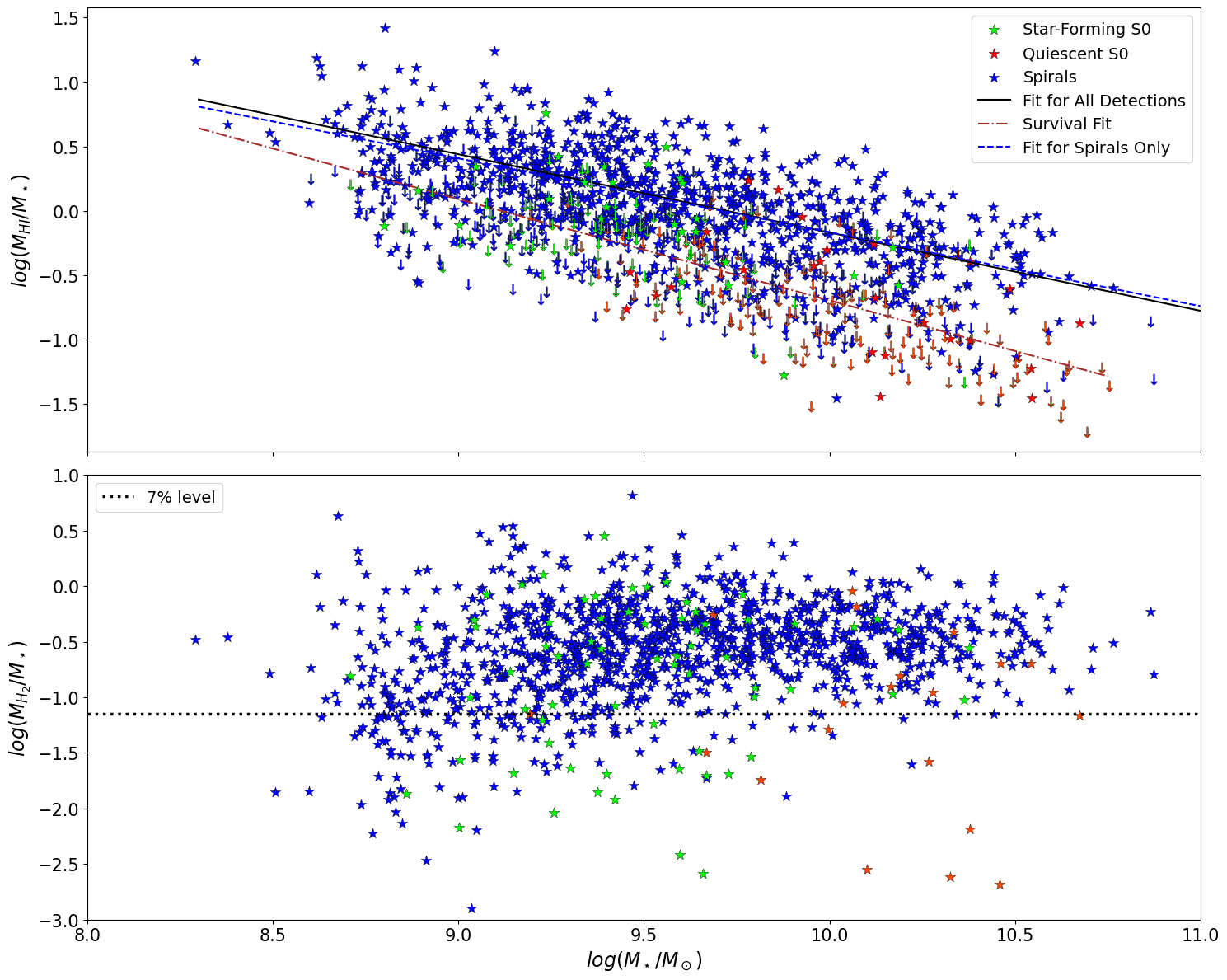}
     \caption{Upper panel: A plot of HI mass fraction as a function of stellar mass for the galaxies used in this sample. As in previous plots we show star-forming spirals (blue; detections as stats, upper limits as down-arrows), with a fit to spiral detections of $\log (M_{\rm HI}/M_\star) = -0.574 \log(M_{\star}/M_\odot)+5.572$ (dashed blue line). We show  both star-forming S0s (green) and quiescent S0s (red/orange). We show a fit of $ \log (M_{\rm HI}/M_\star) =  -0.61 \log(M_{\star}/M_\odot)+5.91 $ (solid black line) which is a fit for all detections in HI-MaNGA (DR4). We also show a survival analysis fit to the entire HI-MaNGA (DR4) sample of $\log (M_{\rm HI}/M_\star) =  -0.786 \log(M_{\star}/M_\odot)+7.166$ (dot-dashed line), and this is the relation used to define HI deficiency in this work (see Section 2.2). Note that the survival analysis line stops at $\log(M_\star/M_\odot) = 10.75$ since the fraction of upper limits in HI-MaNGA DR4 exceeds 50 percent at that point. Lower panel: As above but for molecular gas mass fraction. 
     The markers are the same for both panels. There are five galaxies with molecular gas fractions lower than -0.3 which are not shown.  The horizontal line shows a 7\% mass fraction, which we use as an average value in calculating baryonic masses (Eqn. 3). The mean molecular gas mass is found to be $\log M_{\rm mol}/M_\odot = 8.56$, 8.79 and 8.98 for the SFS0, qS0 and SFSp respectively.}
    \label{fig:3}
\end{figure*}

\begin{figure*}
    \centering
    \includegraphics[width=0.58\linewidth]{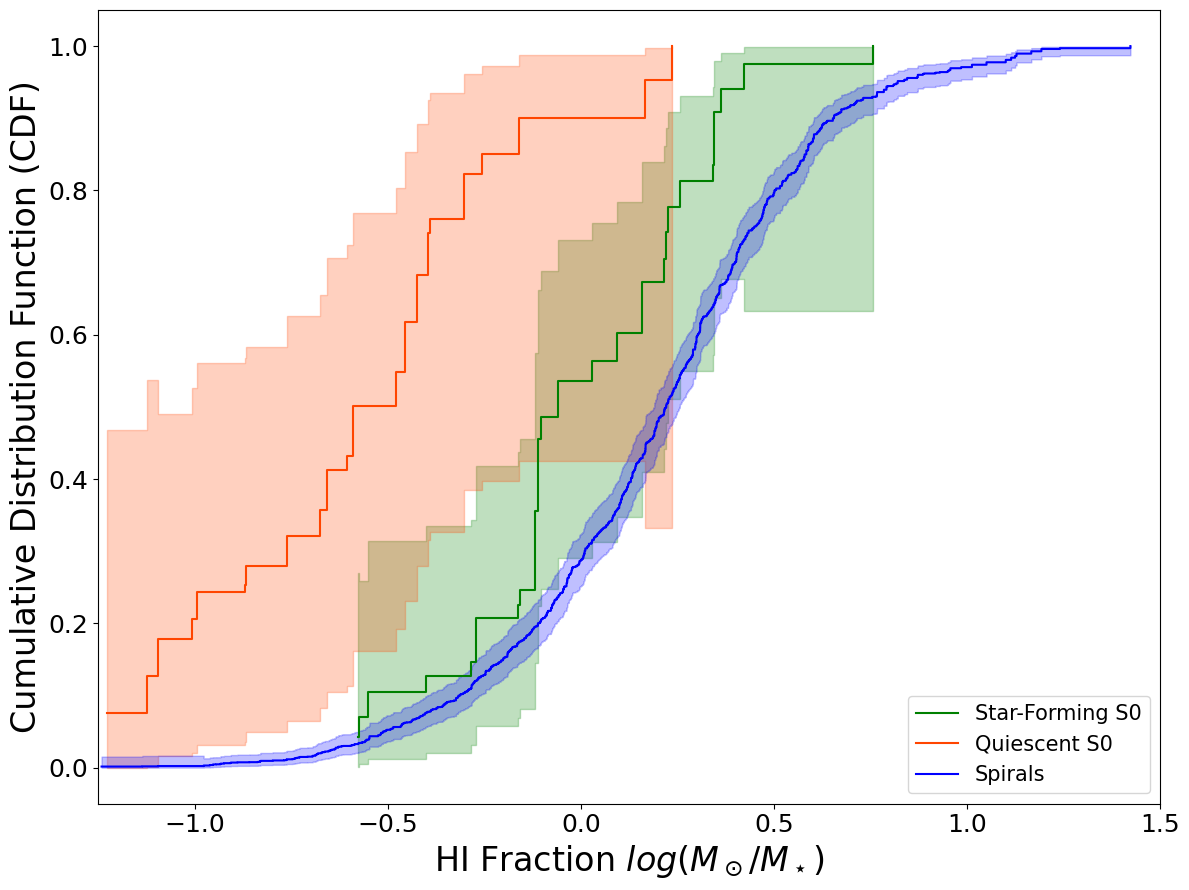} 
    \includegraphics[width=0.58\linewidth]{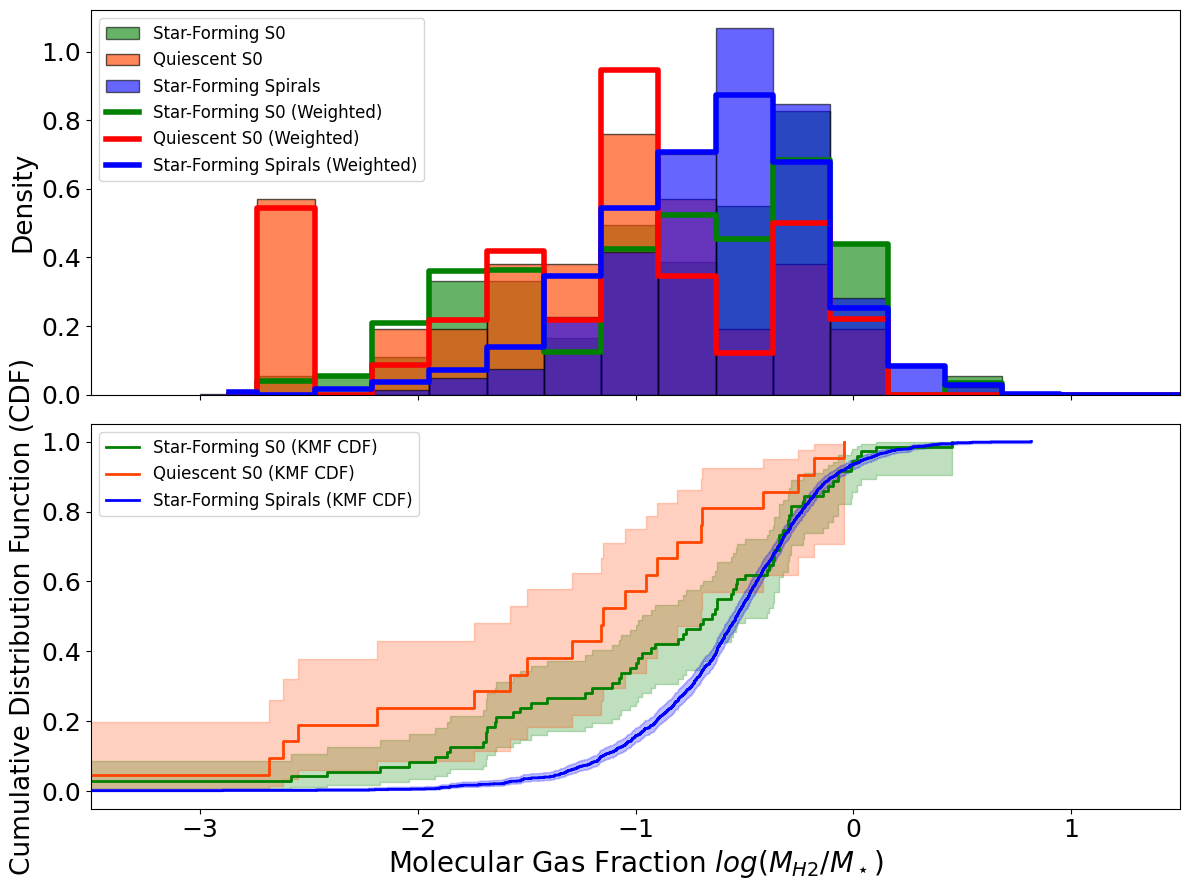}
    \caption{Upper Panel: We show the CDF of HI-to stellar mass ratio (or HI mass fraction), for the three sub-categories of galaxies considered in this work. The median points for the three CDFs are $\log(M_{\rm HI}/M_\star) = -0.6 ^{+0.8}_{-0.9}$ for the quiescent S0s (orange), $\log(M_{\rm HI}/M_\star) = -0.06 ^{+0.2}_{-0.06}$ for the star-forming S0s (green) and $\log(M_{\rm HI}/M_\star) = 0.3^{+0.04}_{-0.03}$ for the star-forming spirals (blue). Comparing distributions we find $p<$ 0.005 between the star-forming S0s and the quiescent S0s (i.e. statistically differing) and $p=0.2$ between the star-forming S0 and the star-forming spirals (i.e. statistically the same). Middle panel: A histogram of the estimated molecular gas fraction, $\log M_{\rm mol}/M_\star$ for the three galaxy populations, along with the CDF plot. Bottom panel: The CDF plot for the molecular gas fraction. The median molecular fraction for the SFS0 is $\log M_{\rm mol}/M_\star = -0.6^{+1.0}_{-0.4}$, for qS0 is $\log M_{\rm mol}/M_\star = -1.2^{+1.6}_{-0.7}$ and for SFSp is $\log M_{\rm mol}/M_\star = -0.5^{+0.6}_{-0.3}$. The solid bars of the histogram show the HI deficiency without the weights while the outlines show the distribution if the weights are applied.}
    \label{fig:4}
\end{figure*}

\subsection{Comparing HI Deficiency/Richness}
	
\begin{figure*}
    \centering
	\includegraphics[height=9cm]{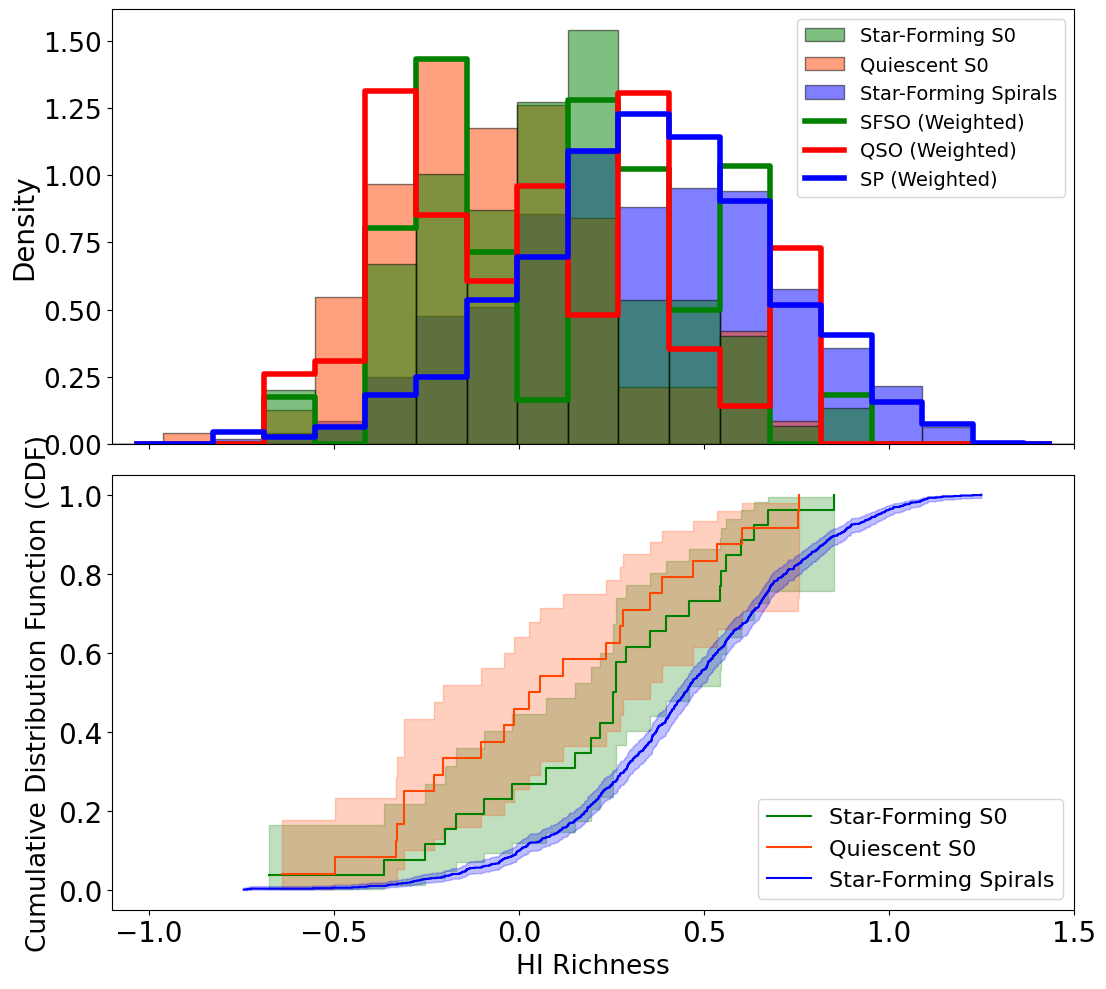}
    \caption{Upper panel: Histograms showing HI richness for the three samples (galaxies with upper limits are included at their upper limit, hence they could be more HI deficient than shown). The solid bars of the histogram show the HI deficiency for the raw sample while the outlines show the distribution if the volume weights (See Section \ref{sec:sample}) are applied. Lower panel: Survival CDF plot showing HI ``richness" (HI$_{\rm rich}=-1.0\times$HI$_{\rm def}$) for the three samples. The medians from the CDFs are HI$_{\rm rich}=-0.6 ^{+0.1}_{-0.2} $ for the quiescent S0, HI$_{\rm rich}= 0.2 ^{+0.04}_{-0.05}$ for the star-forming spirals and HI$_{\rm rich}=- 0.4  ^{+0.1}_{-0.2}$ for the SFS0s. The logrank test reveals $p<0.005$ for comparisons between both the SFS0s and qS0s and the SFS0s and SFSps, showing they are not statistically similar. For comparison with a more limited stellar mass range with maximum overlap between the samples, see Figure \ref{Appendix:1} in the Appendix of this paper.}
    \label{fig:5}
\end{figure*}

As should be clear from Figure \ref{fig:3}, the HI mass fraction strongly correlates with stellar mass. It is also notable that the SF and quiescent S0 samples we make use of in this work have quite different stellar mass ranges (as was previously mentioned in R22). The spiral sample has a mass range which covers both the SFS0 and qS0 range, with a median comparable to the SFS0s. 

Figure \ref{fig:5} shows the distribution of HI deficiency for the three samples both as histograms (right panel) and a CDF (left panel). The histogram includes both detections and upper limits (at the value of the upper limit), and is provided for both the raw sample counts (filled) and the counts weighted by {\tt esweight} to reproduce an effective volume limited sample. We make the CDF for HI$_{\rm rich} = -1\times$ HI$_{\rm def}$, calling this parameter "HI richness". As expected, the highest HI deficiency (lowest "richness") is shown by the quiescent S0's. The SFS0's occupy an intermediate range of HI deficiency, while the star-forming spirals are the least HI deficient (most HI rich). The median values for the HI richness are $\langle$HI$_{\rm rich}\rangle = $ 0.4 $^{+0.1}_{-0.2}$, -0.6 $^{+0.1}_{-0.2}$ and 0.2  $^{+0.04}_{-0.05}$for SFS0, qSO and SFSp samples respectively. While the SFS0s and SFSps have similar median values of HI deficiency for detections, we are able to exclude the hypothesis that any samples have statistically similar distributions of HI deficiencies once upper limits are included. From the CDF we find $p<$ 0.005 for comparison of both the star-forming S0s and the quiescent S0, and the star-forming S0 and the star-forming spirals.  However, as is clear in Figure \ref{fig:1}, the stellar mass distributions of these samples differ significantly, with only a narrow range where all have substantial overlap (i.e. $\log M_\star/M_\odot = 9.5-10$). We investigate this overlap sample in Appendix A. Figure \ref{Appendix:1} should be compared to Figure \ref{fig:5}, showing in this overlap mass range, the qualitative trends we see are similar - although in this smaller sample, both S0 samples have similar HI deficiency, while the SFSps remain richer in HI content.

\subsection{Comparing Gas Depletion Time}

We next look at the gas depletion times for galaxies in the three samples. We will do this for HI and an estimate of molecular H$_2$. 

\begin{figure*}
	\includegraphics[height=8cm]{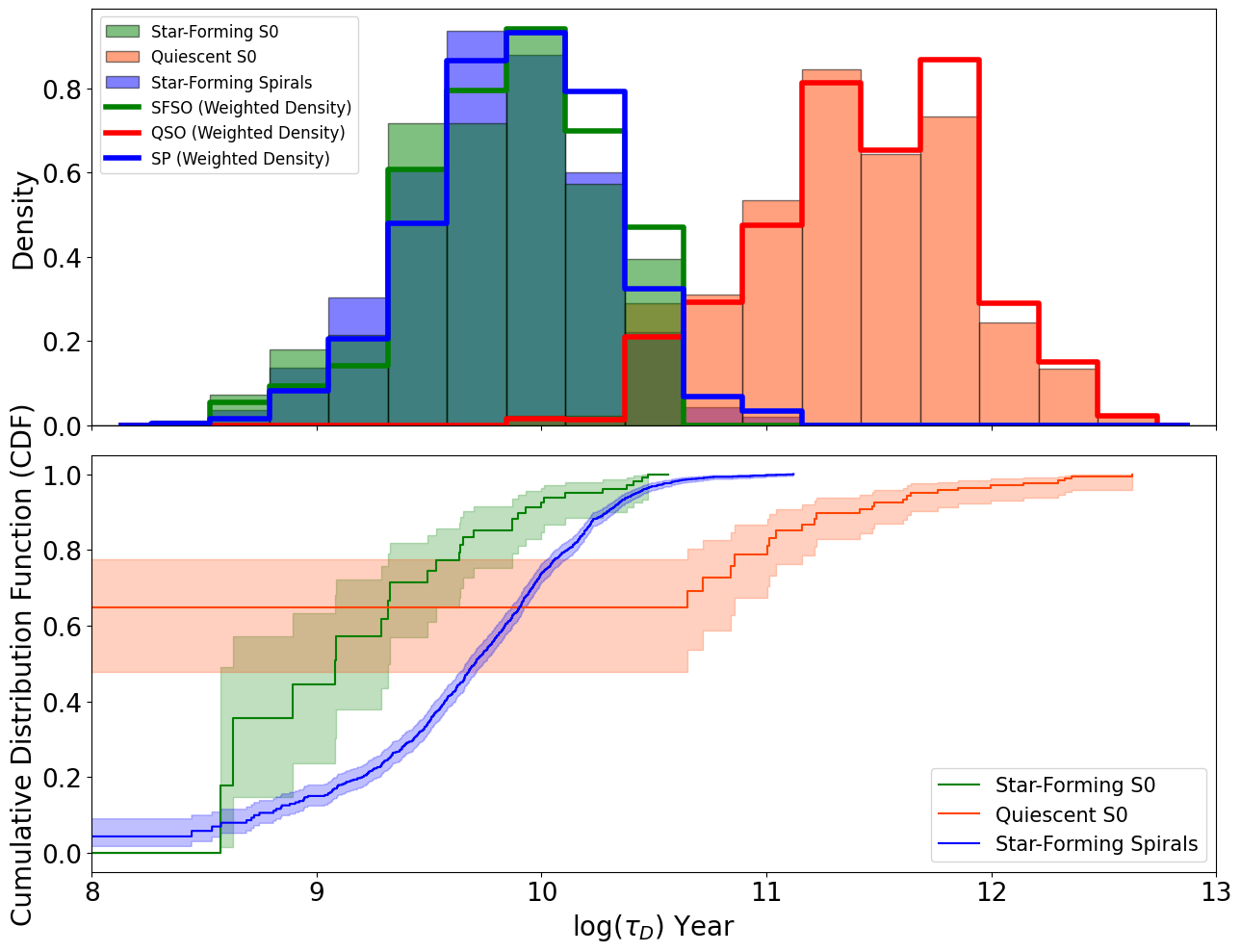}
    \caption { Upper panel: Histograms showing the HI depletion times $\log(\tau_D)$ in years, for the three samples.  This histogram includes both detections and upper limits (included at the value of the upper limit on depletion time). Like in Figure \ref{fig:5},the solid bars of the histogram show the HI deficiency without the weights while the outlines show the distribution if the weights are applied. Lower panel: CDF HI depletion times using survival analysis is shown for the three samples (this is made with the weights taken into account). Star forming S0s have the shortest depletion times of all the samples. The median of the CDF plots are a depletion time of 4.5 Gyrs for the star-forming S0, 9.8 Gyrs years for the star-forming spirals and 302.8 Gyrs for the qS0s. The logrank test shows that all three distributions are statistically different ($p<$ 0.005 between SFS0 and qS0s and $p<0.01$ between SFS0 and SFSp samples). For a limited stellar mass range, a similar plot is included as Figure \ref{Appendix:2} in the Appendix section. }
    \label{fig:6}
\end{figure*}

The distribution of HI depletion times (See Equation \ref{eq:2}) is shown in Figure \ref{fig:6}. As before we show both histograms of the values (with upper limits plotted at their upper limit value) and the CDF using survival analysis techniques. It is notable that quiescent S0s have significantly longer HI depletion times than either SFS0s or spirals; even though they have significantly less HI, their SFR is so much lower that the HI content they have can sustain their current SFR for a long time. By contrast, the SFS0s have depletion times comparable or slightly shorter than the star-forming spirals. The SFS0s are less HI rich than the spirals, so their lower HI means they will run out of the hydrogen reservoir for SF on comparable timescales or slightly more quickly. 

The median depletion times (from the CDF) are found to be of 4.5 Gyrs for the star-forming S0 and 9.8 Gyrs for the star-forming spirals and 302.8 Gyrs for the qS0s. Using the logrank test it is found to be unlikely that the SFS0 depletion times are drawn from the same distribution as either the qS0s ($p<$ 0.005) or the SFSp samples ($p<0.01$). To check the dependence on stellar mass range, we made a similar plot limiting only to the stellar mass range with significant overlap in the samples. This is shown in Figure \ref{Appendix:2}, and demonstrates the same qualitative behaviour is seen in this overlap mass range.

\subsubsection{Molecular Gas Depletion Times} \label{sec:molecular}

The depletion time considered so far has been for the atomic gas content. While HI provides the bulk of the hydrogen mass reservoir for SF, and correlates well with sustained SF \citep[e.g. see][]{Stark_2021}, stars required molecular hydrogen to actually form. At the present time molecular hydrogen measurements are not available for the majority of this sample, however we can look at an estimate of the molecular gas mass of the galaxies derived from the Pipe3D analysis of MaNGA data. These estimates are limited to molecular gas in the region of the galaxy covered by the MaNGA bundle, so may miss extended $H_2$. As noted in \citet{Saintonge2012}  
$H_2$ is typically well mixed with SF (i.e. the optical disc) in nearby galaxies, meaning that the vast majority of molecular hydrogen should be found inside MaNGA bundles (which reach out to 1.5-2.5$r_e$ for 80\% of the MaNGA sample). This was based on observations that CO and SFR distribution trace each other well in nearby star-forming galaxies \citep{Leroy2009AJ}, and even in in the HI-dominated outer disk regions \citep[][]{Schruba2011AJ} and was recently confirmed as a reasonable assumpton by \citet{Salvestrini2025A&A} for 121 star forming galaxies in the ``DustPedia" sample. In Pipe3D, molecular gas mass was calculated from Balmer decrements, which can only be measured in galaxies with H$\alpha$ and H$\beta$ emission lines; much of the S0 galaxy population (particularly the qS0 sample) does not have the necessary data to compute this. For this part of our analysis we are limited to 71 SFS0s (65\% of the sample), 21 quiescent S0s (12\% of the sample), and 1251 star-forming spirals (96\% of the sample), and we note that these (especially the quiescent S0s) are likely biased to include only the most SF. The molecular gas mass fraction is shown for this subset in the lower panel of Figure \ref{fig:3}. As noted in \citet{Sanchez2022}, the Balmer decrement method to estimate dust, assumed gas at $T=10^4$K. We did not explicitly remove galaxies with AGN-like emission in this work, but our sample is based on \citet[][]{Rathore2022} who used selection methods which removed galaxies with broad emission lines, a characteristic of broad-line active galactic nuclei (AGNs). This should remove most AGN contaminated galaxies, although \citet[][]{Rathore2022} do note some narrow line AGN emission may remain.

Figure \ref{fig:7} shows both the estimated molecular gas masses fraction ($\log(M_{\rm mol}/M_\star)$ where available and the molecular gas depletion timescales. It is clear that the SFSp sample has the most molecular hydrogen, followed by the SFS0 sample, and the qS0s have the least molecular hydrogen. The median molecular fraction for the SFS0 is $\log M_{\rm mol}/M_\star = -0.6^{+1.0}_{-0.4}$, for qS0 is $\log M_{\rm mol}/M_\star = -1.2^{+1.6}_{-0.7}$ and for SFSp is $\log M_{\rm mol}/M_\star = -0.5^{+0.6}_{-0.3}$. The median molecular depletion times are 1.3 Gyr, 1.9 Gyr and 171.7 Gyr for the SFS0s, SFSps and qS0s  respectively. From this we can conclude that the shorter HI depletion times for the SFS0s are not made up for by longer molecular depletion times, as they still have the shortest molecular depletion time.  

We note here that missed (extended) molecular hydrogen makes all of these estimates upper limits of the molecular depletion time, but the bias towards only the most SF in each sample (particularly the qS0s, but also the SFS0s) would skew them to lower limits. We do not have the information needed to determine which bias is larger.

\begin{figure*}
    \includegraphics[height=7.6cm]{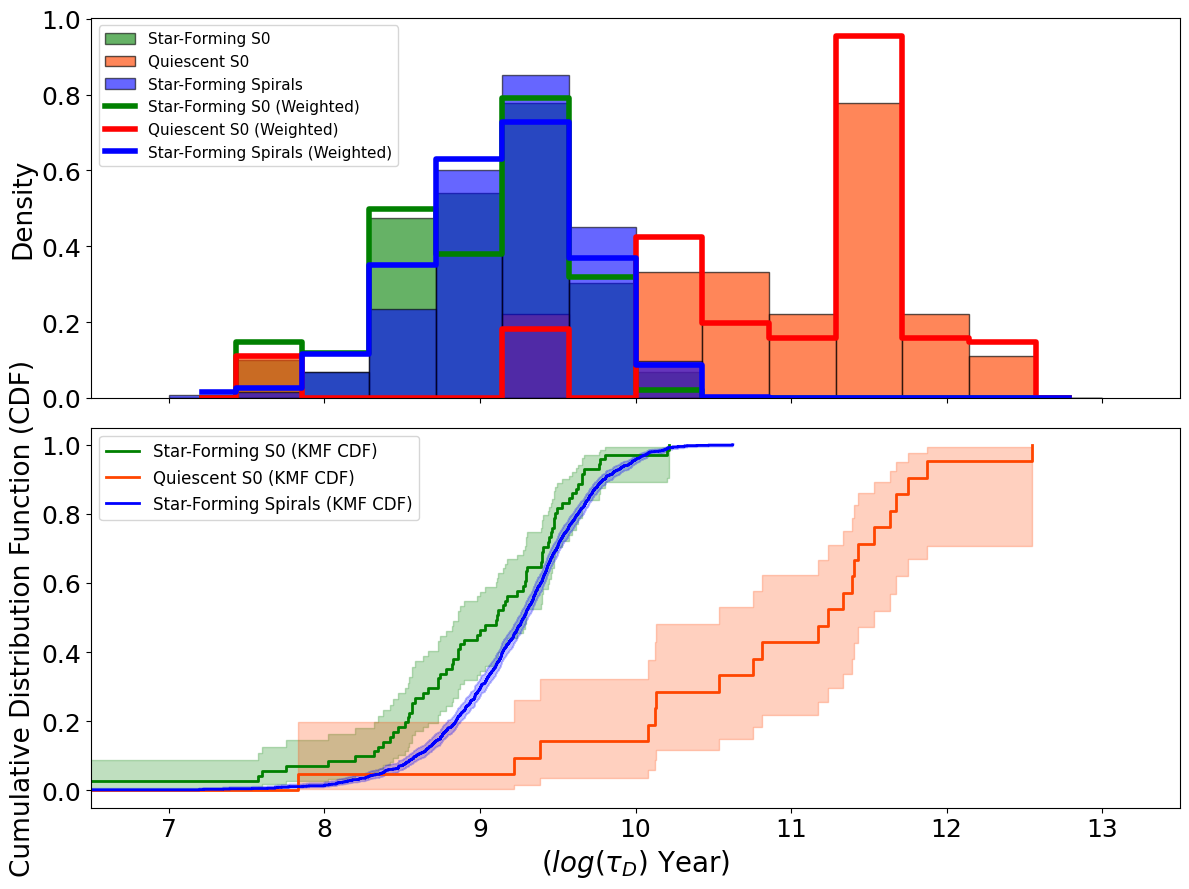}     
    \caption{ Upper panel: The histogram of the estimated molecular gas depletion time for our three galaxy populations. Lower panel: The CDF plot of the molecular gas depletion time. The median depletion times are 1.3 Gyrs for the SFS0, 171.7 Gyrs for the qS0 and 1.9 Gyrs for the SFSp. } 
    \label{fig:7}
\end{figure*}

\subsection{Comparing Baryonic Mass}

\begin{figure*}
	\includegraphics[width=0.6\linewidth]{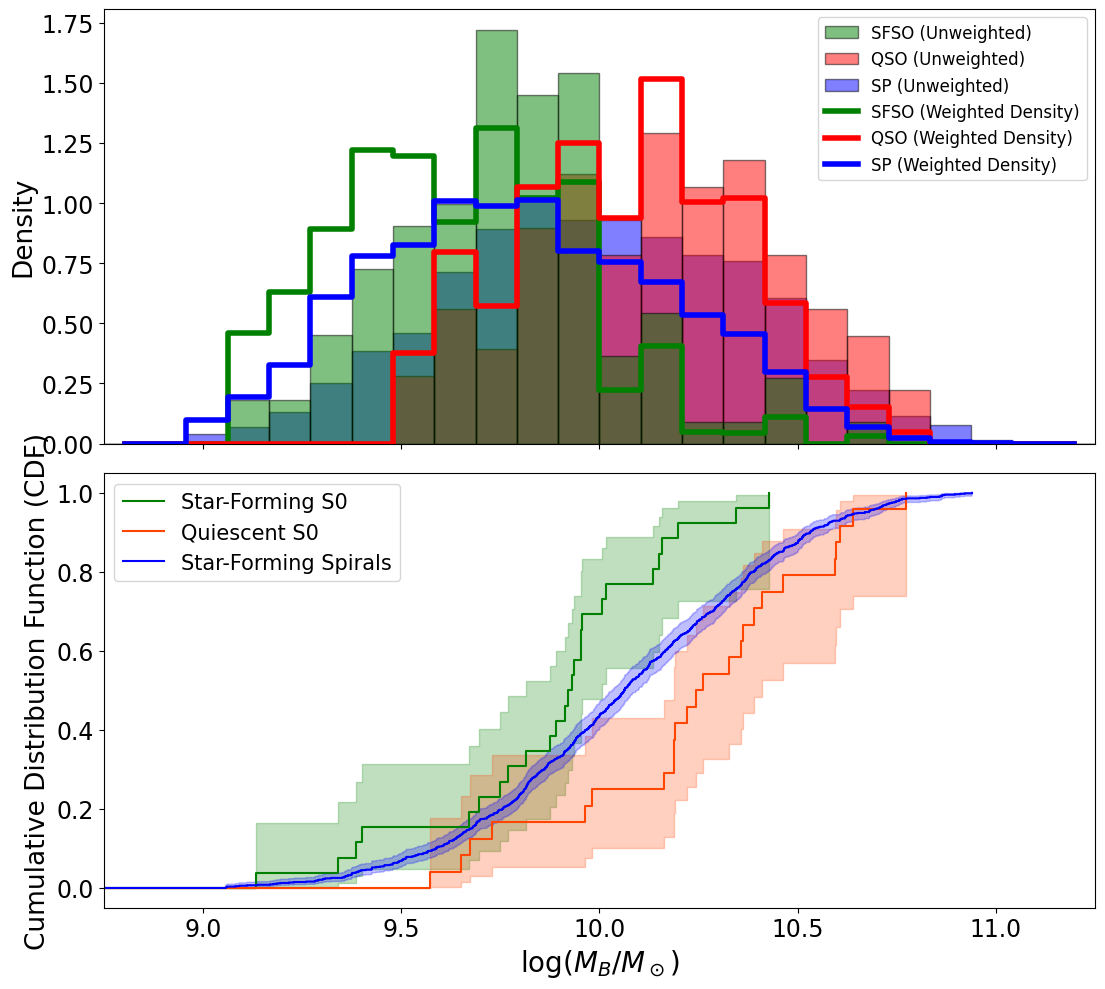}
    \caption{Upper panel: The baryonic mass of the all the galaxies. As in previous plots, the solid bars of the histogram show the Baryonic mass without the weights while the outlines show the distribution if the weights are applied. Lower panel: The Survival CDF plot for the baryonic mass of the galaxies. The median point for the Star-forming S0 sample is  $\log(M_B/M_\odot = 10.4 ^{+1.0}_{-0.04} $,  $\log(M_B/M_\odot = 10.9 ^{+0.6}_{-0.5}$  for the quiescent S0 and $\log(M_B/M_\odot = 10.5 ^{+0.05}_{-0.03}$ for the star-forming spirals. The $p$-value between the star-forming S0 and quiescent S0 samples is $p<$ 0.005 (they are not the same) while the $p$-value between the star-forming S0 and the star-forming spirals is $p=0.08$ (they could be the same). }
    \label{fig:8}
\end{figure*}

Finally, in Figure \ref{fig:8} we investigate the total baryonic mass distribution of our three subsets. Similar to the stellar mass distributions (see R22) we notice that the SFS0s are typically lower baryonic mass than the spirals, while the qS0s in this sample are found at higher masses. Of course, the baryonic masses for HI non-detections are also upper limits, so we also show the CDF using the survival analysis technique. This confirms how different the baryonic mass distributions are for the SFS0s and the qS0s in our sample. We find the median baryonic mass for the SFS0 is
$\log(M_B/M_\odot) = 10.4 ^{+1.0}_{-0.04} $, $\log(M_B/M_\odot) = 10.9 ^{+0.6}_{-0.5}$  for the qS0s and $\log(M_B/M_\odot) = 10.5 ^{+0.05}_{-0.03}$ for the star-forming spirals. We rule out the hypothesis that the baryonic masses of the SFS0s and qS0s are drawn from the same population $p<  0.005$ meaning that the SFS0s cannot turn into the qS0s in this sample by converting gas into stars. However there is marginal evidence that the baryonic masses of the SFS0s and spirals are similar ($p= 0.08$). 

\section{Discussion}

 In summary our findings based on the HI content of a sample of SFS0s compared to both SFSps and qS0s are: 
 \begin{itemize}
     \item SFS0s have marginally lower HI mass fractions and higher HI deficiency than the SFSp samples, but significantly higher fractions and lower deficiency than the qS0s
     \item The HI depletion times of the SFS0s are much shorter than the qS0s, and more similar to, but still shorter than the SFSps. Where we can estimate molecular depletion times we find a similar trend; the SFS0s also have shorter molecular hydrogen depletion times than the qS0s
     \item SFS0s in this sample have lower stellar and baryonic mass than the qS0s. So the SFS0s in this sample cannot turn into the qS0s in this sample by merely turning all their gas into stars. They will still be too low mass.  
\end{itemize}

 We also investigate how the populations differ at fixed stellar mass, picking a range of $\log M_\star/M_\odot = 9.5$-10 where all three samples have significant coverage (see Figures 1 and 3).  This selection removes mostly higher mass qS0s, spirals from both higher and lower mass ranges, and mostly the lower mass SFS0s. The sample sizes get significantly smaller when we do this, with a sample of 72 (41.4\%) qS0s, 394 (30.5\%) spirals and 36 (33\%) SFS0s. Re-running our analysis (see Appendix A for quantitative results), we find that as expected, this moves the HI mass fractions of the three subsets to be more similar, however the differences observed in HI richness and HI depletion times between the subsets are qualitatively the same. In this fixed stellar mass range we also find that the samples have similar total baryonic mass, suggesting the difference in baryonic mass distribution between the qS0s and SFS0s is driven by the difference in stellar mass distribution. This selection results in a sample with molecular gas estimates which is very small, so no conclusions can be drawn about this. 
 
 In the rest of this section, we will compare our observations with other work, and published ideas about the origins of SFS0s. 

\subsection{Building on the work of Rathore et al. 2022}

Our work uses the SFS0, qS0 and SFSp samples constructed by \citet[][, R22]{Rathore2022} from the MaNGA sample. In that work the authors found that
\begin{itemize}
    \item the SFS0s they select have lower stellar mass than quiescent S0s, with most SFS0s having stellar masses, $\log M_\star/M_\odot< 10.25$, while most of the equivalent qS0s are more massive than this.
    \item Radial profiles of SF for the sample, reveal that the SFS0s have their star formation more centrally dominated when compared to the disc-dominated star formation in similarly selected spirals. 
    \item from a comparison of the size-mass relation, $B/T$ and velocity dispersions, SFS0s are structurally more similar to qS0s than spirals, and have velocity dispersions and bulge content similar to typical quenched galaxies.
    \item More than 50\% of the SFS0s are kinematically unsettled
\end{itemize}

R22 concluded that since they found SFS0s are structurally dissimilar to spirals, they are not likely faded spirals. The structural similarities to qS0s suggest rejuvenated quiescent S0s as the main origin of the SFS0s. R22 note that the qS0s which are rejuvenated must be lower mass than those in the sample, and are missing from the comparison set likely due to selection effects in MaNGA. They ended their paper by stating “Atomic and molecular gas observations of the SF-S0s will be very helpful in constraining their evolution, since it might provide insights regarding time-scales and efficiency of star-formation in these objects.” In this work we provide the atomic gas observations (and estimates of molecular gas). 

 Overall we find that the SFS0s have somewhat similar HI masses and depletion times to the spiral sample, and significantly more than the qS0s. The two SF populations have statistically identical HI mass fractions, although SFS0s have statistically slightly higher HI deficiencies 
 This finding suggests that, if it weren't for the structural differences, the SFS0s are reasonably consistent with being faded spirals (where the spiral features disappear before the SF ceases). 
 
  We find (see Figure \ref{fig:6}) that the SFS0 galaxies have relatively, but not extremely low HI depletion times, with $\tau_{D,HI} = 4.5$ Gyrs compared to 9.8 Gyrs for the spiral sample. This suggests the SFS0 phase is relatively short-lived, but not an extremely fast process. For a limited subset of the SFS0s we can also estimate molecular gas depletion times, finding a similar pattern (1.3 Gyrs for the SFS0s and 1.9 Gyrs for the SFSps). 

We find (see Figure \ref{fig:8}) that the SFS0s have on average somewhat lower baryonic mass (M$_b$) than the qS0s in this sample. That means that once the gas has been turned into stars, most of the newly quiescent S0s which result from the faded SFS0s will still be lower mass than the qS0 sample in MaNGA. We agree with R22 that if the bulk of SFS0s form from rejuvenated qS0s there must be a population of lower mass quiescent S0s that are not detected by MaNGA, and add that since the total baryonic mass of the SFS0s is significantly smaller than the qS0s in MaNGA, they do not have enough gas mass to turn into these qS0s after forming stars. The qS0s and SFS0s in this sample are statistical consistent with being drawn from different baryonic mass distributions ($p<0.005$ comparing the two subsets) of galaxies although there is some overlap between the two. 
 
 The primary MaNGA selection based on redshift dependent $i$-band magnitude limits introduces a bias against lower mass quiescent galaxies (and higher mass blue galaxies; \citealt{Wake2017} - see Figure 8 in that paper). It is also known that lower mass galaxies are more likely to be blue/SF and higher mass red/quiescent. A ``colour-enhanced" sample was added to MaNGA to try to mitigate the effect of these observational and galaxy population driven biases, but even with this the MaNGA sample favours lower mass blue galaxies and higher mass red galaxies. A targeted survey of lower mass red galaxies will be needed to test if there is a population of qS0s which could be rejuvenated into the SFS0s and/or that the SFS0s could fade into. 

R22 also looked at kinematic asymmetry as an important clue to the formation of SFS0s. They found that kinematically regular SFS0s had higher stellar mass, while disturbed ones were lower mass. R22 did not claim to fully understand the physical reasons for this, but commented that higher mass galaxies require more disturbance to become kinematically unsettled, and weaker interactions are capable of disturbing lower mass galaxies. We checked the CDFs of the HI deficiency of the regular and disturbed SFS0s (as defined by R22) separately, but could find no statistical differences. While the samples are small, so this could be due to statistical uncertainty, we have no evidence to suggest regular and disturbed SFS0s have any difference in their HI deficiency. Taken at face value this suggests that whatever process is causing the disturbance has not (or perhaps not yet) altered the HI content. With strong HI detections ($S/N>6$) is is possible to measure HI asymmetry (asymmetry in a HI global profile can come from a mixture of kinematic asymmetry, and lopsided HI discs), which would be interesting to compare to the optically detected kinematic asymmetry from R22. Unfortunately, only 14 of the 28 HI detections we have for SFS0s have high enough $S/N$ to make this measurement; deeper single-dish observations, or resolved HI observations would be needed to probe this further. 

\subsection{Comparison with Other HI Observations}

 Historically, most resolved HI observations tend to focus on late-type galaxies, which are typically observed to be significantly more gas rich than ETGs \citep[e.g.][]{Roberts1994}. However, there are several previous works that cover early-type galaxies, and particularly blue early-types.  For example, \citet[][]{Noordermeer2005} used the Westerbork Synthesis Radio Telescope (WRST) to observe 68 early-type disk galaxies finding that their HI properties resemble those of spiral galaxies. While the HI surface densities were slightly lower, the mass-diameter relation held, suggesting that the distribution of HI is not fundamentally different. In contrast, \citet[][]{helmboldt2007} carried out HI measurements using GBT on 30 early-type K+A (i.e. recent post-starburst) galaxies, and they noted that these star-forming early types are on average slightly more gas-poor than typical spirals.

 Many early types with cold gas observations reveal a potential for disc regrowth. For example, \citet[][]{Wei2010ApJ} examined the HI gas content of a sample of low-mass, blue-sequence early-type galaxies and found that many possess sufficient gas to significantly grow their stellar disks. Depending on star formation scenarios (either constant or declining), these galaxies could build up stellar mass over relatively short timescales, indicating a capacity for rejuvenation.  \citet[][]{sil2020star} provided a focused case study of the S0 galaxy UGC 5936. Using HI Data from the WSRT archive combined with optical spectroscopy they were able to observe a star-forming ring with solar metallicity, closely aligned with an HI-rich outer disk. \citet[][]{sil2020star} suggest that the smooth, extended HI morphology of UGC 5936 suggests cold gas accretion from a satellite, fueling extended star formation in a stable, laminar fashion, which would provided ideal conditions for disk regrowth. 

 Many works suggest HI in typical (not SF) lenticular galaxies is significantly more extended than in spirals, and at low densities so that it cannot form stars \citep{Oosterloo2007,Sharma2023}.  Notably, \citet{Oosterloo2007} used the Australia Telescope Compact Array (ATCA) to observe gas rich lenticulars identified in the HIPASS survey \citep{Meyer2004}, finding that about 60\% of the sample have HI in a regularly rotating disc. \citet{Sharma2023} suggest this is consistent with a significant fraction of the very HI rich, but not SF galaxies they identified in HI-MaNGA, but resolved HI data has not yet been obtained for this sample. 

 \citet[][]{Wei2010ApJ} discuss the idea that morphological transformation (from blue early-types to later discs) may be possible if all remaining HI can be converted into stars.  Many of the galaxies in their sample required ongoing external gas inflow to sustain star formation on longer times. They speculate that star formation in their sample galaxies is bursty and likely involves externally triggered gas inflows.

 Evidence of minor-mergers or kinematic asymmetry is also common. For example, \citet[][]{Noordermeer2005} note that many of the early-type galaxies they observed in HI exhibited lopsided gas morphologies, hinting at past or ongoing interactions or minor mergers—key mechanisms for introducing or redistributing cold gas, and while \citet{Oosterloo2007} found a significant number of regular HI discs in early-types, around a third instead showed irregular HI morphologies.  
 
 While working on this project, we became aware that \citet{Chen2025} also carried out some studies on SFS0's using a combination of MaNGA data and HI observations. They construct a sample of 134 SFS0 galaxies from the MANGA sample, using methods similar to R22 (both papers use the sample morphological identification, but differ in how SF is characterized). \citet{Chen2025} construct comparison samples of qS0 and both blue and red spirals, while R22 provides qS0 and SFS0 comparisons.  \citet[][]{Chen2025} report on new HI observations of the most highly star-forming of these SFS0 ($N=41$) with the Five-hundred-meter Aperture Spherical Radio Telescope (FAST), adding an additional 10 galaxies meeting the sample selection which had been previously observed by HI-MaNGA \citep{Stark_2021}. In contrast to the R22 SFS0s, \citet{Chen2025} SFS0s skews to higher stellar masses, with most of their HI observations for galaxies with $\log M_\star/M_\odot > 10$. Similar, to us, however, they find that there is no statistical difference between the HI properties of SFS0s and blue spirals. Both SFS0s and red spirals in their sample have relatively low HI mass fractions, but are found at high stellar masses; once corrected for mass they have typical HI content. Our slightly larger SFS0 sample skews to lower masses, presumably driven by the different SF selection. 
 
\subsection{Pathways to the Formation of SFS0 Galaxies}

\citet{Chen2025} proposed four possible pathways for the evolution between all four types of star-forming and quenched spiral and lenticular galaxies in their sample (see their Figure 9). Three of these are relevant specifically to the formation of SFS0s that we consider here. \citet{Rathore2022} comment that there are three basic possible formation pathways for SFS0s (faded spirals, rejuvenated qS0s or some completely independent formation). In this section we will consider how our HI observations support or conflict with various possible scenarios for how SFS0s form.  

In the \citet{Chen2025} scenarios, SFS0s are suggested to evolve via several mechanisms: 
\begin{itemize}
\item qS0s turn directly into SFS0s via external gas accretion or minor mergers
\item Red spirals turn into SFS0s via external gas accretion or minor mergers, losing their spiral characteristics 
\item Normal blue spirals turn into SFS0s via fading spiral arms or quenching
\end{itemize}
In all of these cases, it is assumed that SFS0s will fairly quickly turn into (or back into) qS0s after quenching/when residual SF is exhausted. 

 External gas accretion has previously been invoked to explain enhanced HI content of early type galaxies \citep[e.g. as suggested in a study of 12 SAURON sample galaxies, ][]{Morganti2006MNRAS.371..157M}. Any external injection of gas, for example, in a minor merger event, should lead to a boost of star formation, if the conditions are right for gas to condense and form into stars. This boost of SF should not last long, and over a period of time, any star-forming S0 would return to being a qS0. Our data is consistent with this picture. We find relatively short timescales for HI depletion of the SFS0s, so they should quickly become qS0s. However we note that because of the different baryonic mass distributions, the the majority of SFS0s in our sample cannot fade to be comparable to the sample of qS0s in MaNGA, but must fade to lower mass qS0s, at least within the mass limits of the MaNGA sample.

 One possibility is that SFS0s form from normal blue spirals after they undergo some process by which the spiral arms disappear. Contrary to this picture are the findings of R22 who suggest the SFS0s in this sample are not structurally similar to spirals - they have higher $B/T$, more centrally concentrated SF, and higher velocity dispersions. However, based on the HI properties we do find that they cover similar baryonic mass ranges, and have similar HI content as the spiral sample. Secular processes, perhaps driven by galactic bars, or the spirals themselves, may be able to concentrate the ongoing SF and/or stellar content, while raising the velocity dispersion and causing the spiral arms to fade. Such a process would be consistent with our observations. Bar structures are known to have an important role to play in quenching of disk galaxies. \citet{Masters2012MNRAS}, for example, found that the fraction of disc galaxies with bars was notably lower in the gas-rich population than gas-poor. Similary, \citet[][]{Cheung2013ApJ} suggest their observed trends of lower bar likelihood in galaxies with higher specific SFR was driven by the higher gas fraction of the disk. Their results suggests that bars are not stagnant structures within disk galaxies but are a critical evolutionary driver of their host galaxies in the local universe ($z < 1$). \citet[][]{Gavazzi2015A&A} using an H-$\alpha$ imaging survey of galaxies in the ALFALFA survey concluded that strong bars could contribute significantly to the suppressed star formation observed in the inner parts of galaxies, comparing their observations to simulated galaxies with/without bars. This points towards the impact of the formation of strong bars in galaxies is an important mechanism in regulating the redshift evolution of the sSFR for field main-sequence galaxies. The R22 selection of S0s we base our study on excludes any galaxy with a bar, so if this is the formation mechanism for SFS0s from spirals the bar (and the spiral) must also be disrupted in the process.  

 Based on our observations, it seems likely that the majority of qS0 detected in MaNGA are a completely different population than the SFS0s here. They must be a result of more massive star-forming galaxies (possibly spirals) slowly depleting their gas and transforming it into stars, or mergers involving low mass progenitors. Referring to Figure \ref{fig:8}, it can be seen that a large percentage of the quiescent lenticulars have a baryonic mass in the mass range of $\log M_B/M_\odot=$10.6-11.2, while the star-forming S0s peak in the baryonic mass range of $\log M_B/M_\odot=$10.4-10.6. Thus the data rules out this population of SFS0s turning into this population of qS0s. Referring back to the pathways proposed in \citet[][]{Chen2025}, our baryonic mass measurements suggest that SFS0 galaxies we observe must be formed from either lower-mass spiral galaxies which have faded or lower mass qS0s that have undergone gravitational interactions and accreted gas. 

 It is also possible that our SFS0s could form via wet mergers from gas rich progenitors that are too low mass to be included in the MaNGA sample. If these low-mass SFS0s use up their gas reserves (both HI and molecular) in star-formation activities they would quickly deplete their gas reservoir and transition into qS0 at the lower mass range of our sample of qS0s. However we note that both kinematically regular and disturbed SFS0s in our sample have similar HI content, so if this is the explanation we must be catching them at an early phase of the merger. This is reinforced by taking the total baryonic mass, which shows most of the qS0s are on the higher mass end of the distribution. There is no way most of the SFS0s in our sample can be rejuvenated qS0s unless the qS0s loose most of their baryonic mass.

To test if the SFS0's are formed via a gradual evolution, morphing from spirals to early-type galaxies while retaining some residual SF, or if these are rejuvenated quiescent objects, likely by some merger or gas accretion event, it would be useful to further constrain the timescales of star-formation. Future work should consider how various tracers of star formation timescales, such as measures in a variety of different wavelength regimes, or measures like $\alpha$-enhancement which probe the timescales of supernovae re-enriching the interstellar medium, may help constrain the formation mechanisms of the SFS0s. 

\section{Conclusions}

Our findings are as follows:
\begin{enumerate}
\item  The HI Mass Fraction and HI Deficiency of the star-forming S0 galaxies in our sample lie between the fraction for the quiescent S0 and the star-forming spirals (but are more similar to the SFSps). 
\item The HI depletion time of the quiescent S0s is over 300 Gyrs, significantly longer than that of both the SFS0 (4.5 Gyrs) and spirals (9.8 Gyrs). It is notable that the SFS0s the shortest  depletion times in our sample. While the sample of star-forming S0's in HI-MaNGA do have gas reserves, they cannot sustain star-formation for as long as a typical star forming spiral. The quiescent S0's are by definition, not forming stars, hence even a tiny amount of gas can sustain their low level SF for very long times. 
\item Comparisons of the estimated molecular gas fraction shows that the quiescent S0 also has the lowest fraction of the three populations, with a corresponding molecular depletion time of 171.7 Gyr, followed by the SFS0s with a molecular depletion time of 1.3 Gyr and then the star-forming spirals with a molecular depletion time of 1.9 Gyr. The SFS0s do not appear to have large reserves of molecular gas compared to the SFSps.
\item A comparison of the total baryonic masses of the samples indicate that the SFS0s in the MaNGA sample are significantly lower total baryonic mass than the quiescent S0. Spirals in our sample are found to be more similar to the SFS0s. This finding may be driven by the selection criteria for MaNGA \citep{Wake2017}.
\end{enumerate}

We suggest a possible formation pathway of the SFS0 to be lower-mass progenitors (not detected by MaNGA) which are either low mass QSOs which accrete external gas, or lower mass spirals which are structurally changed to match the SFS0s.

The massive quiescent S0s in MaNGA, despite being structurally similar to the SFS0s (R22), are a completely separate population, with very different mass range. We suggest they are most likely to be formed either from fading star-forming spirals or from medium- to high-mass spirals undergoing mergers. The SFS0s in this sample cannot form from the quiescent S0s in this sample that receive in-falling gas, and the qS0s in this sample cannot form from fading SFS0s. 

The puzzle of the origin of unusual SFS0s remains interesting; our observations of the neutral atomic gas content provide an important clue to possible pathways for their formation from lower-mass progenitors. Future observations with additional time awarded to the HI-MaNGA project will help to further refine the findings of this work; current HI-MaNGA observations are targeting massive galaxies with longer integration times to either reach more stringent HI upper limits, or increase the number of HI detections for massive galaxies. 

This work adds to a growing literature supporting a paradigm shift in our understanding of early-type galaxies. Once considered passively evolving systems, especially at low redshift, many early-type galaxies, particularly those on the blue sequence, harbor substantial HI reservoirs and demonstrate active or potential stellar disk regrowth. The evidence of cold gas accretion, disk-like HI distributions, and ongoing star formation points to transitional or rejuvenating phases in galaxy evolution.  It has long been understood that morphological classification alone is insufficient to determine a galaxy's entire evolutionary state. Instead, external processes (e.g., interactions, mergers, gas accretion) and internal star formation capacity must be considered, along with the impact of internal secular evolution. This work supports the idea that S0s are dynamically evolving systems that can show rejuvenation of star formation at times. In the broader context of galaxy evolution astronomers now have a more complex, dynamic and cyclical view: early-type galaxies are not necessarily evolutionary endpoints with a single formation mechanism, but may form via multiple pathways, and even oscillate between quiescent and star-forming phases depending on their gas content, internal structures and interaction history. 

\section*{Acknowledgments}

The Green Bank Observatory is a facility of the National Science Foundation operated under cooperative agreement by Associated Universities, Inc. This work uses data from proposals GBT16A-095, GBT17A-012, GBT19A-127, GBT20B-033 and GBT21B-130. 

MaNGA is part of Sloan Digital Sky Survey IV. Funding for the Sloan Digital Sky Survey IV was provided by the Alfred P. Sloan Foundation, the U.S. Department of Energy Office of Science, and the Participating Institutions. SDSS-IV acknowledges support and resources from the Center for High Performance Computing at the University of Utah. The SDSS website is www.sdss.org.

Some assistance from Gemini AI (the Google Collab integrated AI) was used to generate the codes for some of the plots in this work. 

KLM and DVS acknowledge support from the NSF under grant AST-2510740 "Collaborative Research: Understanding the extreme diversity of atomic hydrogen depletion times in galaxies". 

\section*{Data Availability}
The HI-MaNGA catalogue (DR4) used in this study can be found at https://greenbankobservatory.org/science/gbt-surveys/hi-manga/. 


\bibliographystyle{mnras}
\bibliography{ReferencePaper1} 


\appendix

\section{Analysis for Mass-Limited Sample}

{In this appendix we show results limiting our sample to a stellar mass range of $\log(M_\star/M_\odot)=9.5-10$, or the mass range where our samples show the greatest overlap (see Figure 1). Table \ref{tab:Table_1} summarizes the sample sizes in this range and gives average values of the HI fraction, HI richness and HI depletion times. Figure \ref{Appendix:1} reproduces Figure 5 from the main paper, showing the distribution of HI richness only in this mass range, while Figure \ref{Appendix:2} is the equivalent of Figure 6 showing the distribution of HI depletion times in this mass range. From these figures we conclude the qualitative results presented in the main text also apply in this narrow range where all subsamples are comparably represented. }

 \begin{table*}
	\centering
	\caption{A summary of the sample size of the different galaxy samples in the mass range of $\log(M_\star/M_\odot)=9.5-10$ only. Values for the entire sample are shown in parentheses for ease of comparison.}
	\label{tab:Table_1}
	\begin{tabular}{lcllccr} 
		\hline
		Category & No. of galaxies & Median HI Fraction & Median HI Richness & Median HI Depletion time (Gyr) \\
		\hline
		Quiescent S0 & 72 (174) & -0.40 (-0.6$^{+0.8}_{-0.9}$) & -0.86 (-0.6$^{+0.1}_{-0.2}$)  & 406.0 (302.8) \\
		Star-Forming S0 & 36 (109) & -0.11 (-0.06$^{+0.2}_{-0.06}$) & -0.68 (+0.4$^{+0.1}_{-0.2}$) & 4.3 (4.5) \\
		Star Forming Spirals & 394 (1291) & +0.02 (+0.3$^{+0.04}_{-0.03}$) & +0.31 (+0.2$^{+0.04}_{-0.05}$)&  6.5 (9.8) \\
		\hline
	  \end{tabular}
\end{table*}

\begin{figure*}
	\includegraphics[width=10cm]{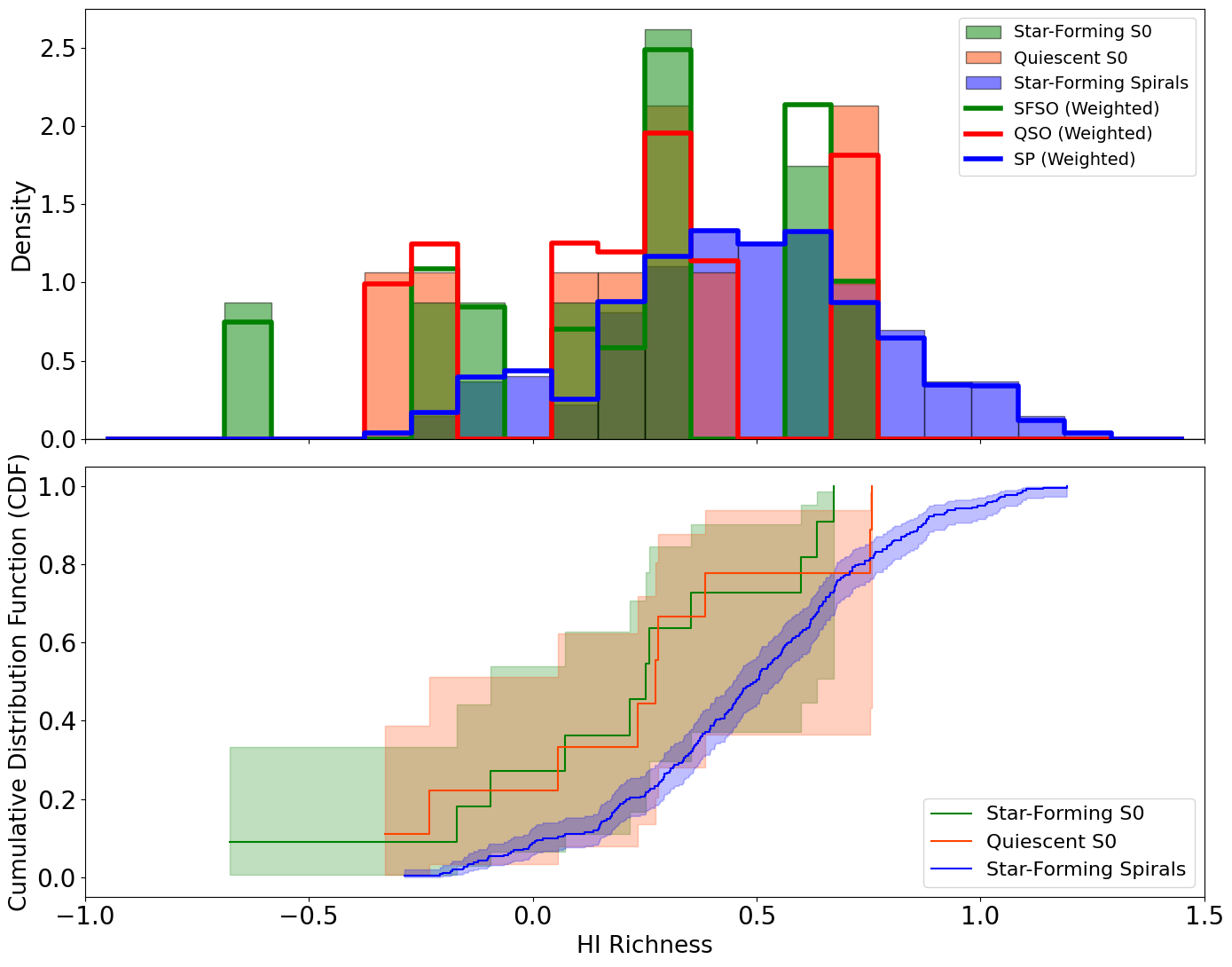}
    \caption{The HI richness histogram and CDF for the stellar mass range of $\log(M_\star/M_\odot)=9.5- 10$. This figure corresponds to Figure \ref{fig:5} in the main text. }
    \label{Appendix:1}
\end{figure*}

\begin{figure*}
	\includegraphics[width=10cm]{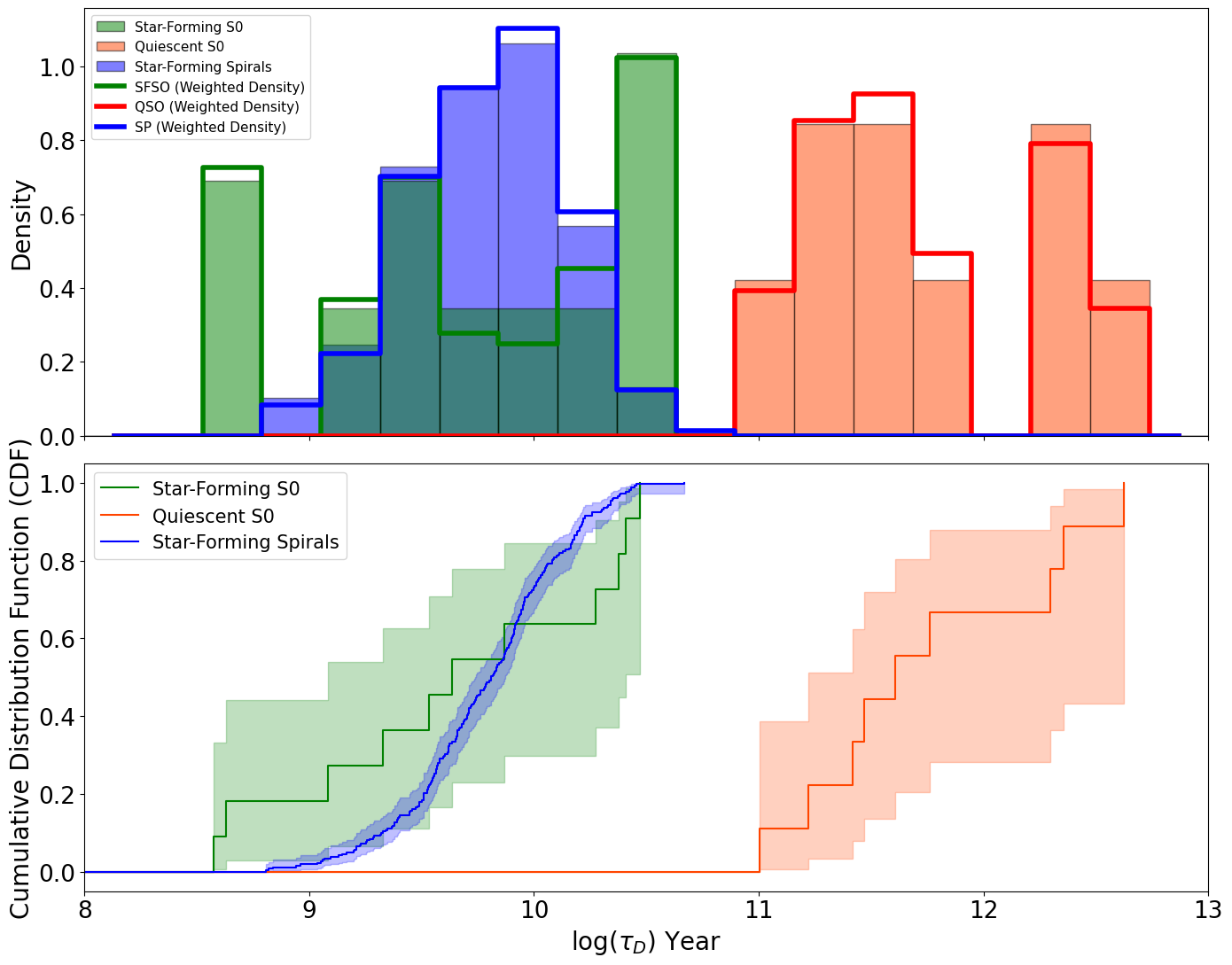}
    \caption{The HI Depletion time histogram and CDF for the stellar mass range of $\log(M_\star/M_\odot)=9.5- 10$. This figure corresponds to Figure \ref{fig:6} in the main text.}
    \label{Appendix:2}
\end{figure*}


\bsp	
\label{lastpage}
\end{document}